\documentclass{osa-article}

\journal{oe}

\newcommand{\ket}[1]{|{#1}\rangle}
\newcommand{\bra}[1]{\langle{#1}|}

\begin{document}

\title{Direct generation of frequency-bin entangled photons via two-period quasi-phase-matched parametric downconversion}

\author{Fumihiro Kaneda\authormark{1,2*}, Hirofumi Suzuki\authormark{1}, Ryosuke Shimizu\authormark{1,3}, and Keiichi Edamatsu\authormark{1}}

\address{\authormark{1}Research Institute of Electrical Communication, Tohoku University, 2-1-1 Katahira, Sendai, 980-8577, Japan\\
\authormark{2}Frontier Research Institute for Interdiciplinary Sciences, Tohoku University, 6-3 Aramaki aza Aoba, Aoba-ku, Sendai 980-8578, Japan\\
\authormark{3}University of Electro-Communications (UEC), 1-5-1 Chofugaoka, Chofu, Tokyo 182-8585, Japan}
\email{\authormark{*}kaneda@riec.tohoku.ac.jp} 



\begin{abstract}
We report a simple scheme for direct generation of frequency-bin entangled photon pairs via spontaneous parametric downconversion. 
Our fabricated nonlinear optical crystal with two different poling periods can simultaneously satisfy two different, spectrally symmetric nondegenerate  quasi-phase-matching conditions, enabling the direct generation of entanglement in two discrete frequency-bin modes. 
Our produced photon pairs exhibited Hong-Ou-Mandel interference with high-visibility beating oscillations--- a signature of two-mode frequency-bin entanglement. 
Moreover, we demonstrate deterministic entanglement-mode conversion from frequency-bin to polarization modes, with which our source can be more versatile for various quantum applications. 
Our scheme can be extended to direct generation of high-dimensional frequency-bin entanglement, and thus will be a key technology for frequency-multiplexed optical quantum information processing. 
\end{abstract}

\bibliographystyle{osajnl.bst}

\section{Introduction}
Photons and their entanglement are key resources for quantum information processing (QIP) including quantum teleportation, quantum computations, and quantum communication. 
In particular, \textit{polarization} entangled photons generated via spontaneous parametric downconversion (SPDC) \cite{Kwiat:1995ub} have been successfully used in many proof-of-principle demonstrations of QIP protocols \cite{Edamatsu:2007fl,Pan:2012kv}.
Meanwhile, SPDC photon pairs often have entanglement in another degree of freedom; due to energy conservation in the SPDC process, downconverted photons can be naturally entangled in \textit{frequency}. 
This frequency entanglement is undesirable in polarization-encoded QIP, in which photons need to be indistinguishable from each other except in polarization states for utilizing two-photon interference \cite{Hong:1987vi}, a key effect in two-qubit quantum gates \cite{Knill:2001is}. 
To meet such requirements spectral engineering techniques \cite{Grice:2001jc,Valencia:2007bq,Mosley:2008ki,Evans:2010jn,YABUNO:2012kk,Kaneda:2016fh,Chen:2017gt} have been developed for eliminating frequency entanglement. 
However, while only two orthogonal states can be used in polarization encoding, a number of modes can be extracted in the frequency domain, and therefore frequency encoding can be efficiently used for high-dimensional QIP applications such as high-capacity quantum communications  \cite{Olislager:2010is,Zhong:2015iu,Mirhosseini:2015fya} and mode-multiplexed photonic state engineering \cite{McCusker:2009ci,Kaneda:2015dn,Kaneda:2017gp}.

In this paper, we consider an entangled state with $n$ discrete frequency modes: 
\begin{equation}
\ket{\textrm{FB}_n} = \sum_{j = 1}^{n} \frac{1}{\sqrt{n}}\ket{\omega_j}\ket{\omega_{n-j+1}}, 
\label{n-mode fbin}
\end{equation}
where $\ket{\omega_j}$ denotes a single-photon state having center frequency $\omega_j$ and a bandwidth much narrower than the peak frequency separation $\omega_{j}-\omega_{j' \neq j}$. 
This ``frequency-bin'' entangled state is more useful than continuous frequency entanglement in the context of QIP that in general requires the manipulation of dicrete modes of photons (although broadband continuous frequency entanglement is also a key resource in some important applications such as clock synchronizations \cite{Giovannetti:489607}, quantum optical coherence tomography \cite{Nasr:2008gu,Okano:2015bf}, and probing nonlinear optical process \cite{Nakatani:2011bi}). 
Experimental observation of two-mode frequency-bin entanglement was first reported by Ou and Mandel in 1988 \cite{Ou:1988tn}; 
however, the behavior of frequency-bin entanglement was extracted from continuous frequency entanglement by postselective projection measurements via two-photon coincidence detections. 
Such postselective schemes cannot directly distribute entanglement over remote parties for quantum communication and networking. 
Similar postselective schemes have been demonstrated in later works \cite{Rarity:1990ux,Shih:1994wb,Li:2009jy,Zhou:2014gc,Zhou:2014fn}. 
Non-postselective, direct generation of two-mode frequency-bin entanglement have been demonstrated by using SPDC with interferometers \cite{Ramelow:2009bs,Jin:2016bm,Jin:2018cj} and noncollinear phase-matching conditions \cite{Kim:2015ks}. 
However, for experimental simplicity, system stability, and compatibility with integrated optics, a single-pass, collinear SPDC is more desirable. 
Recently, Xie \textit{et al.}, \cite{Xie:2015uw} have demonstrated remarkable 17-mode frequency-bin entanglement, but that experiment used a resonator cavity for spectral filtering that may result in limited transmissions and generation rates.

Here, we report a simple scheme for direct generation of two-mode frequency-bin entangled photons at telecom wavelengths. 
A nonlinear optical crystal with two poling periods simultaneously satisfies two different quasi-phase-matching (QPM) conditions \cite{Ueno:2012tc,Herrmann:2013vn}, directly generating frequency-bin entanglement between orthogonally polarized photons. 
Therefore, our scheme does not need either spectral filtering, postselective measurements, or interferometers. 
Our scheme can be extended for generating larger-mode frequency-bin entanglement by introducing additional poling periods for higher-dimensional QIP applications.  
We also demonstrate entanglement-mode conversion from frequency-bin to polarization modes, allowing us to choose an appropriate entanglement degree of freedom for different applications.

\section{Principle of operation}

\begin{tiny}
\begin{figure}[t!]
   \centering\includegraphics[width=0.8\columnwidth ,clip]{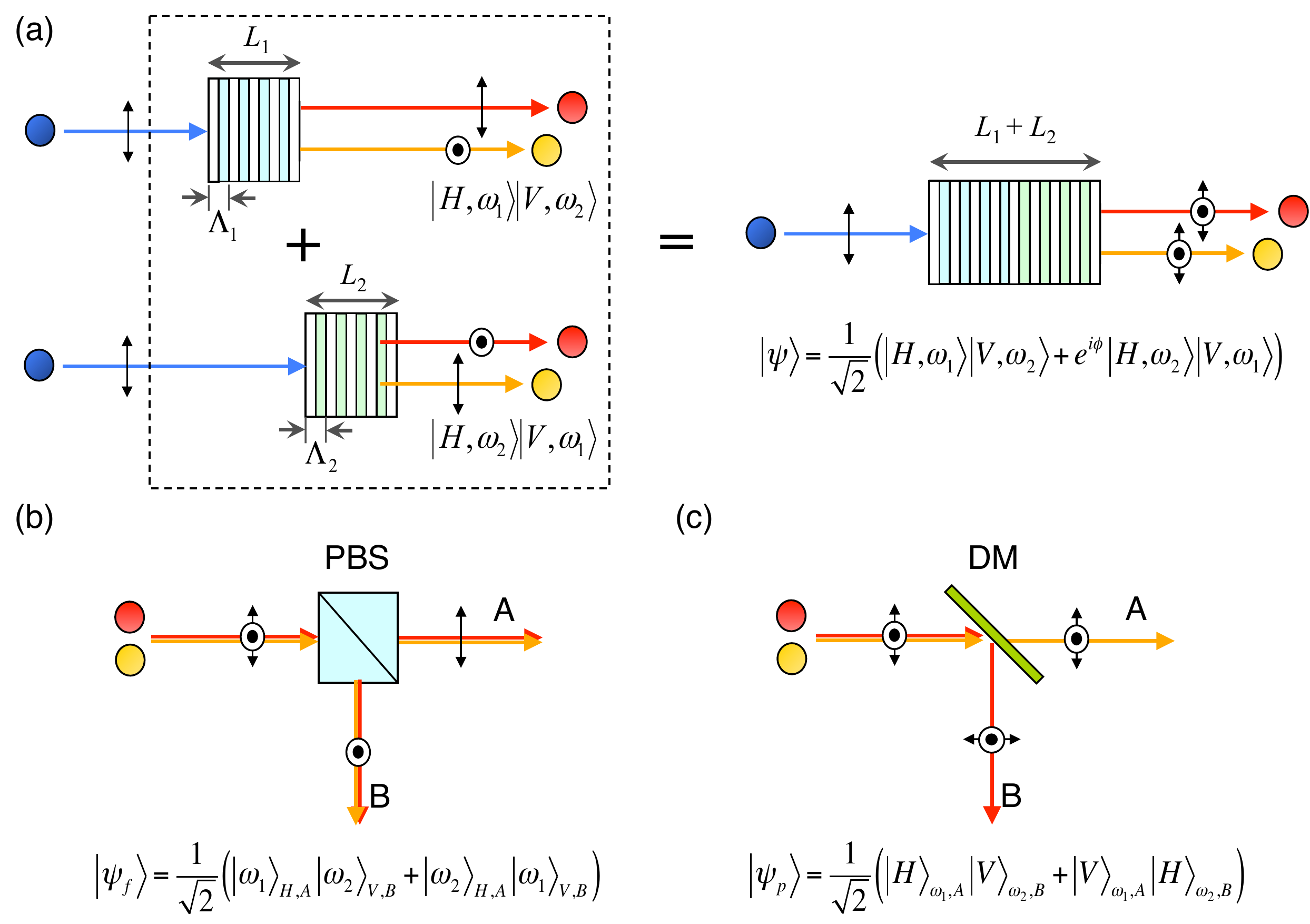} 
   \caption{Schematic diagram of the generation of frequency-bin entangled photon pairs. (a) The illustration of two-period QPM crystal. (b) Frequency-bin entanglement generation. (c) Polarization entanglement generation demonstrated in \cite{Ueno:2012tc}. PBS: polarizing beamsplitter, DM: dichroic mirror. } \label{two period qpm}
\end{figure}
\end{tiny}

Figure \ref{two period qpm} shows a schematic diagram of the frequency-bin entanglement generation. 
A nonlinear crystal has two sequential poling structures with the period $\Lambda_1$ and $\Lambda_2$ over the lengths $L_1$ and $L_2$, respectively (see Fig. \ref{two period qpm}a). 
In general, SPDC has the highest conversion efficiency when the pump and downconverted (signal and idler) modes satisfy energy and momentum conservation
\begin{equation}
 \Delta \omega  =\omega_p - \omega_s -  \omega_i = 0, 
\label{energyconservation}
\end{equation}
\begin{equation}
\Delta k    = k_p- k_s -  k_i - \frac{2\pi} {\Lambda_{1(2)}}= 0, 
\label{phasematching}
\end{equation}
where $\omega_l$ and $k_l$ are angular frequency and the wavenumber of the pump ($l = p$), signal ($l = s$), and idler ($l = i$) modes, respectively. 
In our scheme, the two poling periods are determined to generate othrogonally polarized, spectrally symmetric two-photon states via type-II collinear QPM conditions; 
 $\Lambda_1$ and $\Lambda_2$ are, respectively, designed to produce $\ket{H,\omega_1}\ket{V,\omega_2}$ and $\ket{H,\omega_2}\ket{V,\omega_1}$,  
where $\ket{H(V),\omega_{1(2)}}$ denotes a single-photon state having horizontal (vertical) polarization and frequency $\omega_{1(2)}$. 
Thus, a two-photon state emitted from the two-period crystal can form a superposition state  
\begin{align}
\ket{\psi} = \frac{1}{\sqrt{2}}(\ket{H,\omega_1}\ket{V,\omega_2} + e^{i\phi}\ket{V,\omega_1}\ket{H,\omega_2}),
\end{align}
where $\phi$ denotes the relative phase: $\phi = 2 \pi(1/\Lambda_1- 1/\Lambda_2)L_2$ after the two-period crystal, but this phase can be manipulated by introducing birefringence and/or dispersive media \cite{Ueno:2012tc}, as will be demonstrated.   
By spatially separating photons in the state $\ket{\psi}$ in terms of polarization, e.g., with a polarizing beamsplitter (PBS), one can obtain a two-mode frequency-bin entangled state, 
\begin{align}
\ket{\psi_f} = \frac{1}{\sqrt{2}}(\ket{\omega_1}_{A,H}\ket{\omega_2}_{B,V}+e^{i\phi} \ket{\omega_2}_{A,H}\ket{\omega_1}_{B,V}),
\label{freq_entanglement}
\end{align}
where $A$ and $B$ represent the spatial modes of photons. 
Thus, our scheme can directly generate frequency-bin entanglement via single-pass collinear SPDC without any spectral filtering or postselective measurements. 
Moreover, our scheme can be extended to the generation of arbitrary $n$-mode frequency-bin entanglement with a crystal having $n$ different poling periods in which $\Lambda_i$ produces $\ket{H,\omega_i}\ket{V,\omega_{n-i+1}}$. 
As has been reported in \cite{Ueno:2012tc} and as shown in Fig. \ref{two period qpm}c, by splitting collinear photons in $\ket{\psi}$ with a dichroic mirror (DM), a nondegenerate polarization entangled state can be obtained: 
\begin{align}
\ket{\psi_p} = \frac{1}{\sqrt{2}}(\ket{H}_{A,\omega_1}\ket{V}_{B,\omega_2}+e^{i\phi} \ket{V}_{A,\omega_1}\ket{H}_{B,\omega_2}).
\label{pol_entanglement}
\end{align}
Note that $\ket{\psi_f}$($\ket{\psi_p}$) can be deterministically changed back to $\ket{\psi}$ by recoupling two photons with a PBS(DM): 
therefore, as will be demonstrated, the entanglement in frequency mode can also be deterministically transferred to polarization mode, and vice versa.

\section{Experiment}

\begin{tiny}
\begin{figure}[t!]
     \centering\includegraphics[width=0.9\columnwidth,clip ]{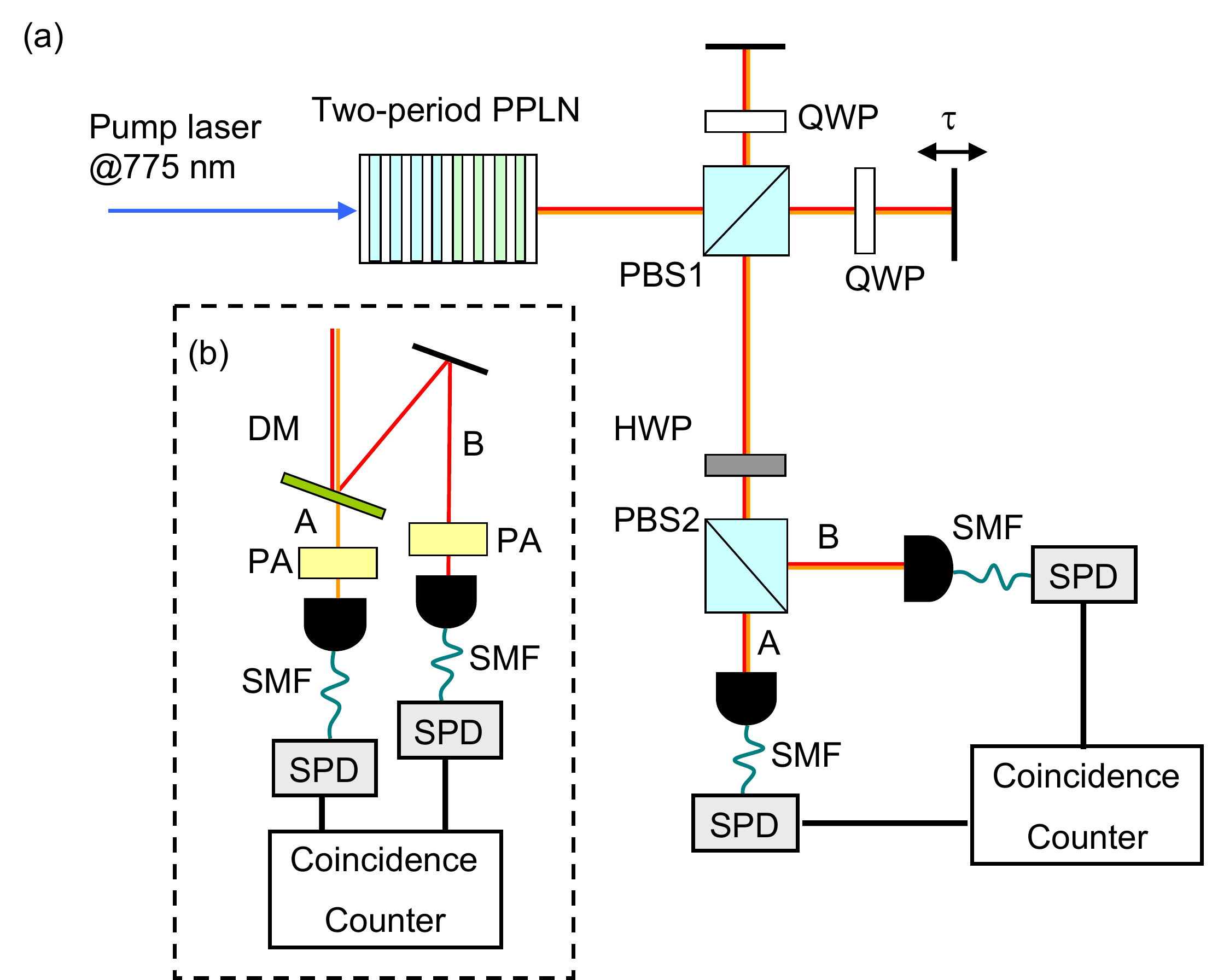} 
   \caption{Illustration of the experimental setup for (a) generation and detection of frequency-bin entangled photons and (b) detection of polarization entanglement. PPLN: periodically poled lithium niobate, PBS: polarizing beamsplitter, QWP: quarter-wave plate, HWP: half-wave plate, SMF: single-mode fiber, SPD: single-photon detector, DM: dichroic mirror, PA: polarization analyzer.}\label{setup}
\end{figure}
\end{tiny}

Our experimental setup for generation and characterization of frequency-bin entangled photons is illustrated in Fig. \ref{setup}a. 
We designed and fabricated \cite{Ueno:2012tc} a periodically-poled LiNbO$_3$ (PPLN) crystal with $\Lambda_1$ = 9.25 $\mu$m and $\Lambda_2$ = 9.50 $\mu$m and $L_1,L_2 = L \sim$20 mm, so that two different QPM conditions have the same photon-pair generation probability. 
The crystal is pumped by second harmonic (wavelength of 775 nm) of an amplified external cavity diode laser with a repetition rate of 4 MHz and a  pump pulse width of 2.5 ns, much longer than the LN crystal length to eliminate the temporal distinguishability of produced photons.   
The crystal temperature was kept at 120.0 $^\circ$C with an accuracy of 0.01 $^\circ$C where the two different QPM conditions generated the same peak wavelengths at 1506 nm and 1594 nm with a $\sim$1.4 nm bandwidth. 

The frequency entanglement of the produced photons was characterized by Hong-Ou-Mandel interference (HOMI) \cite{Hong:1987vi}:
In general, a photon pair produced in distinguishable spectral modes (e.g., $\ket{\omega_1}_{A}\ket{\omega_2}_{B}$) does not exhibit HOMI (unless the coincidence detection time window is much shorter than the reciprocal of the frequency detuning $\delta \omega = \omega_1 -  \omega_2$, as demonstrated in \cite{Zhao:2014cf}). 
However, despite having distinguishable spectral modes, frequency-bin entangled photons, an equal coherent superposition of $\ket{\omega_1}_{A}\ket{\omega_2}_{B}$ and $\ket{\omega_2}_{A}\ket{\omega_1}_{B}$, are perfectly indistinguishable as a two-photon joint state, and therefore can exhibit HOMI. 
In other words, HOMI experiments can be a direct test of entanglement of photons generated in discrete frequency modes. 

Our experimental setup can measure HOMI for photons in two orthogonal polarization modes rather than spatial modes used in conventional HOMI experiments \cite{Hong:1987vi}. 
In a polarization-mode Michelson interferometer, our collinear SPDC photons are first split by PBS1 and thereby converted to $\ket{\psi_f}$ in Eq. (\ref{freq_entanglement}). 
After being reflected by mirrors the photons come back to PBS1 with a time delay $\tau$ between the two spatial modes of frequency-bin entangled photons. 
This variable time delay $\tau$ introduces a phase $\phi = \delta \omega \tau$ to see HOMI over the two-photon coherence length, as well as to compensate a group delay between $H$- and $V$-polarized photons \cite{Ueno:2012tc}. 
A half-wave plate (HWP) rotated at 22.5$^\circ$ and $\text{PBS}2$ work together as a 50:50 beam splitter for $H$/$V$ polarization modes. 
The photons coupled into single-mode fibers (SMFs) are detected by a pair of single photon detectors  (SPD, ID Quantique id201). 
Coincidence events of the two SPDs are recorded by a time interval analyzer (EG\&G 9308) with a coincidence window of 2.5 ns. 
In our experimental setup, typical single and coincidence count rates are $1.7 \times 10^4$/s and $8.5 \times 10^2 $/s, respectively. 
With those single and coincidence rates, we estimate that the average system transmission (including SPDs' $\sim$25\% and $\sim$18\% detection efficiencies at 1506 nm and 1594 nm) is $\sim$5\%. 

\section{Frequency-bin entanglement}
\begin{tiny}
\begin{figure}[t!]
\centering   \includegraphics[width=0.8\columnwidth , clip]{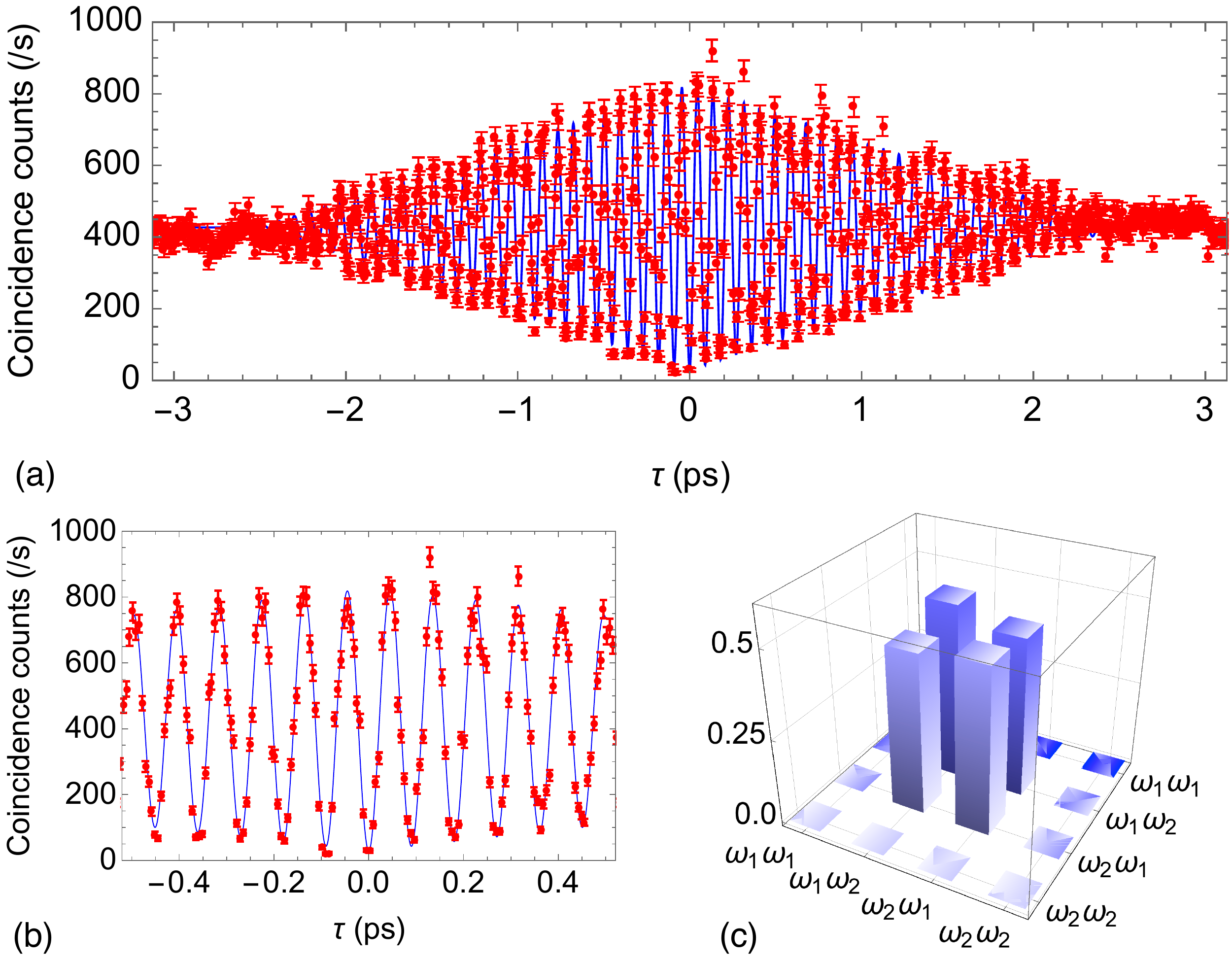}   
\caption{Characterization of the frequency-bin entanglment. (a) Observed HOM interference for (a) -3 ps$\leq \tau \leq $ 3 ps and  (b) -0.5 ps$\leq \tau \leq $ 0.5 ps. (c) Reconstructed density matrix. Error bars were calculated from Poissonian photon-counting statistics.  }
\label{result1}
\end{figure}
\end{tiny}   

Figure \ref{result1}a shows the result of HOMI for our produced photons and Fig. \ref{result1}b is a close-up view of Fig. \ref{result1}a around $\tau = 0$. 
Here, we set the origin ($\tau = 0$) where we observed the lowest coincidence rate. 
The observed high-visibility sinusoidal oscillations are a clear signature of the frequency-bin entanglement. 
The solid curve is the best fit to the experimental data with a fitting function \cite{Ramelow:2009bs}
\begin{align}
I(\tau) = 
\begin{cases}
\frac{N}{2}\{ 1- V\cos (\delta \omega \tau ) (1-|{\textstyle \frac{\tau}{\tau_c}}|)\}  &\quad  \textrm{for}\quad |\tau| \leq \tau_c \\
\frac{N}{2}  &\quad  \textrm{for}\quad |\tau| > \tau_c 
\end{cases}
\end{align}
where the $N$ and $V$ represent the coincidence count rate and interference visibility, respectively. 
$\tau_c$ denotes the full width at half maximum of the triangular envelope predicted from a convolution of two-photon rectangular temporal amplitudes \cite{Rubin:1994vc}. 
We obtained $\delta \omega /2\pi$ = 11.5$\pm $ 0.5 THz and $\tau_c = 2.40 \pm 0.03$ ps (and corresponding single-photon bandwidth of $1.47 \pm 0.11$ nm) in excellent agreement with our single-photon spectral measurements. 
Our observed interference visibility is $V = 93.4 \pm 1.0\%$, which is mainly limited by the distinguishability of the two photon wave packets due to slightly different peak amplitudes and group delays in the two different QPM conditions; our predicted upper limit accounting for the different group delays is $V =$ 95.4\% \cite{Ueno:2012tc}. 
We note, however, that a perfect indistinguishability can be achieved by adjusting the ratio $L_1/L_2$ and poling duty cycles \cite{Chen:2017gt} to control group delays and effective nonlinearities in a LN crystal, respectively. 
The degradation of visibility may also arise from slight misalignment of the Michelson interferometer, multi-photon-pair emissions, and dark counts in the SPDs.

In order to quantify the entanglement in our produced state, we estimated its density matrix in two-frequency-bin space. 
By the method introduced in \cite{Ramelow:2009bs} a frequency-bin mode density matrix can be parametrized as 
\begin{align}
\rho_F = p \ket{\omega_1}\ket{\omega_2}\bra{\omega_1}\bra{\omega_2}_{AB} + 
(1-p) \ket{\omega_2}\ket{\omega_1}\bra{\omega_1}\bra{\omega_2}_{AB} \notag \\
+\frac{V}{2}(e^{i\phi}\ket{\omega_1}\ket{\omega_2}\bra{\omega_2}\bra{\omega_1}_{AB}
+e^{-i\phi}\ket{\omega_2}\ket{\omega_1}\bra{\omega_1}\bra{\omega_2}_{AB} ), 
\end{align}
where $p$ ($0 \leq p \leq 1$) represents the probability of $\ket{\omega_1}_A\ket{\omega_2}_{B}$. 
Here we assume that frequency correlated modes (i.e., $\ket{\omega_1}_A\ket{\omega_1}_B$ or $\ket{\omega_2}_A\ket{\omega_2}_B$) have no amplitudes, recalled by the energy conservation in Eq. (1). 
Thus, we simply obtained $p =0.516\pm0.012$ from the ratio of the coincidence count rates of $\ket{\omega_1}_A\ket{\omega_2}_B$ and $\ket{\omega_2}_A\ket{\omega_1}_B$ with a setup in Fig. \ref{setup}b that can perform spectral projection measurements by a DM. 
Additionally, using the phase tunability by the Michelson interferometer, we selected $\phi = 0$. 
Our reconstructed density matrix is shown in Fig. \ref{result1}c. 
The fidelity to the maximally entangled state $\ket{\psi_f}$ (for $\phi = 0$) and the concurrence of $\rho_F$ are $F = 96.7 \pm 0.1\%$ and $C = 0.934 \pm 0.005$, respectively. 
We note that this highly entangled state was obtained without spectral filtering.

\begin{tiny}
\begin{figure}[t!]
   \centering\includegraphics[width=0.9\columnwidth,clip ]{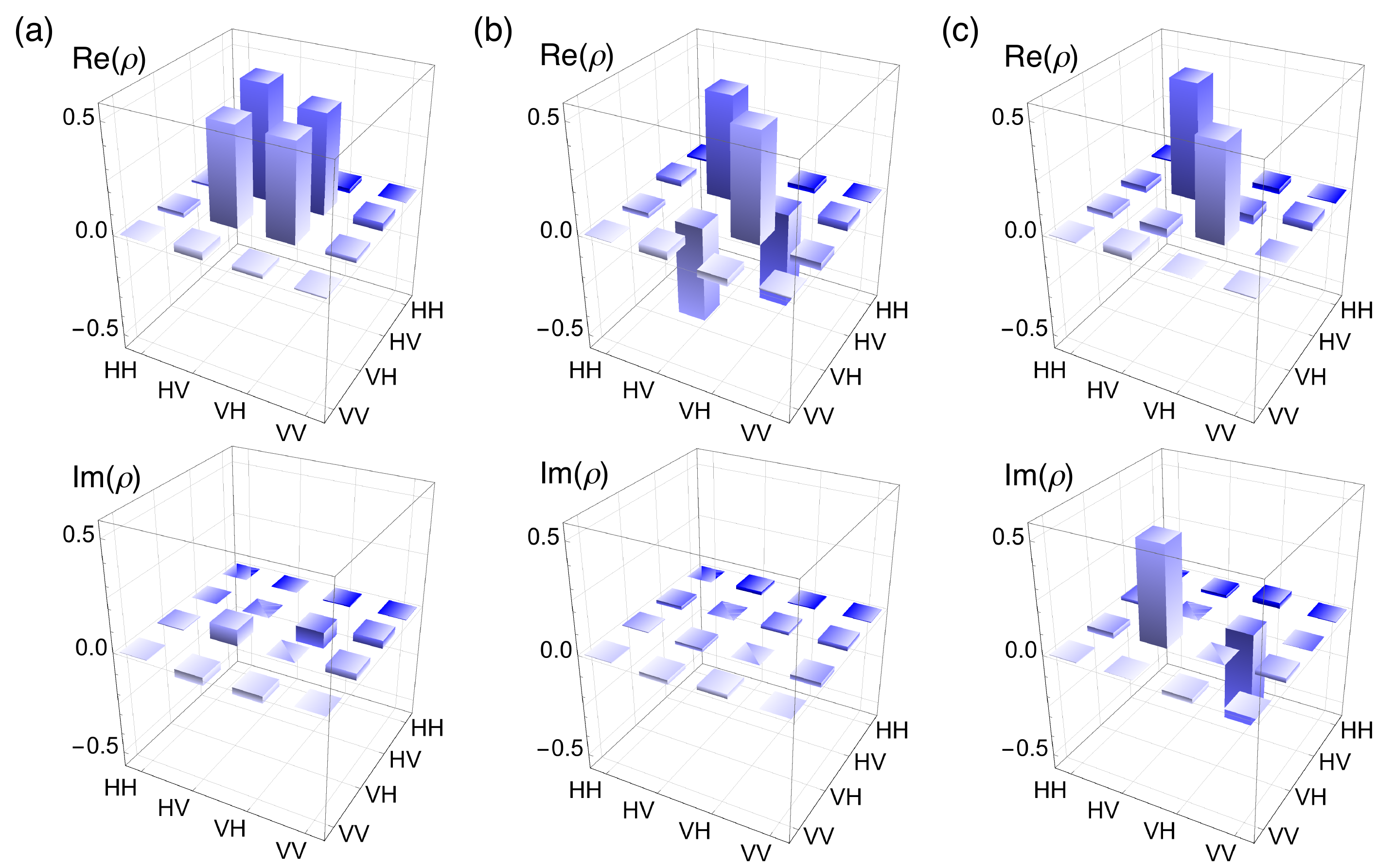} 
   \label{result2}
   \caption{Measured polarization-mode density matrices after the entanglement-mode transfer for $\tau$ = (a) 0 fs, (b) 47 fs, and (c) -20 fs. }
\end{figure}
\end{tiny}

\section{Entanglement mode conversion and symmetry of photon pairs }
The entanglement in our produced photons can be deterministically transferred from frequency-bin to polarization domains. 
Figure \ref{setup}b illustrates our setup (a similar setup has been used for direct generation of polarization entanglement \cite{Ueno:2012tc}). 
As described above, after the creation and phase adjustment of the frequency-bin entangled state, the Michelson interferometer transmits the photons into the same output port of PBS1. 
The collinear photons are then split by the DM with respect to the photon's frequency, deterministically converted to nondegenerate polarization entangled photons $\ket{\psi_p}$ as shown in Eq. (\ref{pol_entanglement}). 

Figure 4a,b,c show our reconstructed density matrices via maximum likelihood quantum state tomography \cite{James:2001bb} for $\tau = $ 0 fs, 47 fs, and -20 fs, where the HOMI in Fig. 3b shows dark, bright, and intermediate fringes, respectively. 
While all of the three density matrices have large populations in $\ket{H}_A\ket{V}_B$ and $\ket{V}_A\ket{H}_B$, phases of off-diagonal elements clearly change with $\tau$. 
Using these results we investigate the relations of the HOMI and the phase of polarization entangled states. 
In HOMI experiments using a non-polarizing 50:50 beamsplitter \cite{Hong:1987vi} two-photon states behave differently due to the different symmetry of the spatial wave functions \cite{Wang:2006dq}; spatially (anti-)symmetric two-photon states are spatially (anti-)bunched after the beamsplitter. 
For an entangled two-photon state, such spatial (anti-)symmetry can be indirectly introduced by (anti-)symmetrized states in an entanglement degree of freedom, since the overall wave function needs to be symmetric in bosonic systems \cite{Fedrizzi:2009hp}. 
In this context, our HOMI experiment using a polarization-mode splitter (shown in Fig. \ref{setup}a) can be interpreted as a test of the polarization-mode symmetry, which can be introduced by the phase of the frequency-bin entanglement adjusted by $\tau$. 
Table \ref{pol_vs_HOMI} summarizes characteristics of the reconstructed density matrices together with the normalized coincidence rates and phases of the frequency-bin entanglement measured by our HOMI experiment. 
For $\tau = $ 0 fs, where we observed two-photon bunching (i.e., the lowest coincidence rate), the polarization state has a high fidelity ($F = 95.6$\%) to the symmetric state (i.e., $\phi = 0$), while the polarization state is closely matched (with $F = 95.9$\%) to the anti-symmetric state ($\phi = \pi$) for $\tau = $ 47 fs, where anti-bunching (i.e., the highest coincidence rate) was observed. 
Those observed relations are consistent with our predictions discussed above. 
For the three different time delays within a beating oscillation period, we obtained comparable values of the fidelities ($F \sim $ 95\%) and concurrences ($C \sim $ 0.93) to those of the frequency-bin entangled state, since the envelope of the HOMI is $\sim$27 times wider than the beating oscillation period.  

\begin{table}[t!]
\begin{center}
\caption{Characteristics of the polarization-mode density matrices. $\tau$, time delay in the Michelson interferometer; $I(\tau)/N$, normalized coincidence count rate in the HOMI; $\phi$, phase of the frequency entangled state predicted from the HOMI; $F$, state fidelity to the ideal density matrix $\ket{\psi_p}$ for $\phi$; $C$, concurrence. }
\label{pol_vs_HOMI}
\begin{tabular}{lccc}\hline
$\tau$    & $0$ fs & 47 fs & -20 fs  \\  \hline
$I(\tau)/N$ & $0.034\pm0.006$ & 0.972$\pm0.033$ & 0.496$\pm0.024$ \\
$\phi = \delta \omega \tau$  & 0 & $\pi$ &1.5$\pi$  \\
 $F$    & 95.6 $\pm$0.9 \% & 95.9$\pm$1.0 \% & 95.6$\pm$0.3 \% \\
$C$ & 0.932 $\pm$0.018  & 0.935$\pm$0.018  & 0.924$\pm$0.006 \\ \hline
\end{tabular}
\end{center}
\end{table}

\section{Conclusion}
In conclusion, we have demonstrated the direct generation of frequency-bin entangled photons, using nondegenerate, two-period QPM-SPDC. 
Our proposed scheme has produced high-quality frequency-bin entanglement with a collinear, single-pass SPDC, not requiring spectral filters or postselective measurements. 
Frequency-bin entanglement of our produced photons was confirmed by measuring high-visibility beating oscillations in HOMI. 
Moreover, we demonstrated deterministic entanglement-mode conversion from frequency-bin to polarization modes, and 
verified the correlation of the symmetry of polarization states and the bunching and anti-bunching properties of the HOMI for the frequency-bin entangled photons. 
Our scheme can be extended to the generation of entanglement in $n$ frequency-bin modes by periodically poling a QPM crystal with $n$ different periods.
Such high-dimensional frequency entanglement will be a key resource for high capacity quantum communications as well as for frequency-multiplexed QIP applications. 
Our multi-period QPM scheme is also compatible with using waveguide structures to enhance the generation rates for practical QIP applications. 

\section*{Funding}
Grant-in-Aid for Creative Scientific Research (No. 17GS1204) from the Japan Society for the Promotion of Science (JSPS).

\section*{Acknowledgement}
We thank Mark Sadgrove and So-Young Baek for helpful discussions. 


\end{document}